\documentclass[12pt]{report}
\usepackage[letterpaper]{geometry}
 \usepackage[doublespacing]{setspace}
 \usepackage{tocloft}
 \usepackage[rm, tiny,center, compact]{titlesec}
 \usepackage{indentfirst}
 \usepackage{etoolbox}
\usepackage{tocvsec2}
 \usepackage[titletoc]{appendix}
 \usepackage{appendix}
\usepackage{rotating}
\usepackage[ruled,vlined,linesnumbered]{TemplateFiles/algorithm2e}
\usepackage{amsthm}
\usepackage{amsmath}
\usepackage{mathrsfs}
\usepackage{graphicx}
\usepackage{amsfonts}
\usepackage{amsmath}
\usepackage{amssymb}

\DeclareFontFamily{OT1}{pzc}{}
\DeclareFontShape{OT1}{pzc}{m}{it}{<-> s * [1.10] pzcmi7t}{}
\DeclareMathAlphabet{\mathpzc}{OT1}{pzc}{m}{it}
\DeclareMathOperator*{\argmin}{argmin}  
\usepackage{xifthen}

\newtheorem{prop}{Proposition}
\newtheorem{remark}{Remark}
\usepackage{pgfplots} 
\newlength\figureheight 
\newlength\figurewidth
\pgfplotsset{compat=newest} 
\pgfplotsset{plot coordinates/math parser=false} 

\allowdisplaybreaks[1]

\usepackage{algpseudocode}
\usepackage{microtype}
\usepackage{lipsum}

\title{\LARGE \bf Decentralized State Estimation via a Hybrid of Consensus and Covariance intersection}
\author {Amirhossein Tamjidi, Suman Chakravorty, Dylan Shell}

\begin{document}
	
	\maketitle
\begin{abstract}
This paper presents a new recursive information consensus filter for decentralized dynamic-state estimation.  No structure is assumed about the topology of the network and local estimators are assumed to have access only to local information. The network need not be connected at all times. Consensus over priors which might become correlated is performed through Covariance Intersection (CI) and consensus over new information is handled using weights based on a Metropolis Hastings Markov Chains. We establish bounds for estimation performance and show that our method produces unbiased conservative estimates that are better than CI. The performance of the proposed method is evaluated and compared with competing algorithms on an atmospheric dispersion problem.   
\end{abstract}

\section{Introduction}
This paper studies decentralized estimation using multiple robotic agents with applications to the estimation of a dynamic random field. When the field dynamics can be described by a linear, lumped-parameter model, the classical solution is the Kalman filter (KF). However, bandwidth and energy constraints may preclude the centralized implementation of such a filter and necessitate the design of a decentralized estimator. 

In general, a decentralized sensor network cannot achieve the estimation quality of a centralized estimator but is inherently more flexible and robust to network failure and consequently is advantageous in certain applications \cite{Zhang_ttradeof1}.

In decentralized estimation settings, the system comprises a set of nodes connected to each other through a communication network with some topology. Nodes are assumed to make noisy observations of a global state from which the full state of the system cannot necessarily be recovered. The goal is to design local estimators that can recursively calculate an estimate of the global state with access only to the information locally available to nodes. We desire that estimates be conservative and the estimator be consistent. No prior knowledge about the network topology is assumed. 

When the topology of the network is known \textit{a priori} and it remains connected throughout, some existing methods recover the centralized estimator's performance \cite{olfati2005distributed,olfati2005consensus} for dynamic state estimation. However, such methods are not applicable for the case where the network does not remain connected all the time.


For static state/parameter estimation Xia, et al., introduce a method based on distributed averaging that can converge to the global state estimator provided that the infinitely occurring communication graphs are jointly connected \cite{Boyd2005}. This method relies on the distributed averaging property of Metropolis-Hastings Markov Chains (MHMC). The advantage of it is that the network topology can be dynamically changing and it need not be connected at all times. The local estimators exchange information only with their immediate neighbors and remain agnostic about the topology of the rest of the network. Their work is limited to static field/parameter estimation. In the dynamic state estimation, when the network becomes disconnected, the estimate priors can drift away while they still have some mutual information. Performing distributed averaging on those priors is wrong since it results in multiple counting of the mutual information. In order to solve this problem one would have to resort to decentralized estimators that account for the correlations between local estimates.  

In \cite{capitan_delayed_icra_2009}, a Decentralized Delayed-State Extended Information Filter (DDSEIF) is described that handles the correlation between local estimates. This method only works in directed networks that do not have any loops. It is claimed that under certain assumptions local estimates would converge to the centralized estimate. However, the method requires a large amount of data communication, storage memory, and book-keeping overhead, and therefore, does not lend itself to online resource constrained recursive distributed state estimation.

Another approach to deal with the correlation of local estimates is to use Covariance Intersection (CI) methods \cite{chen2002estimation} that produce conservative estimates in the absence of correlation knowledge. The work in references \cite{wang_distr_CI,chen2002estimation, dist_inf_filter_2009,hu2012diffusion,deng2012sequential,cristofaro2013distributed} fall into this category. They propose different optimization criteria to perform CI and/or use different iterative CI schemes for decentralized state estimation. 

The downside of decentralized CI based methods is that they produce overly conservative estimates by unnecessarily performing the covariance intersection on generally uncorrelated new information at the current step. This incurs significant performance loss compared to MHMC based distributed averaging, which is a superior way to reach consensus on uncorrelated information.

\begin{figure*}[!b]
	\centering
	{\includegraphics[width=16cm]{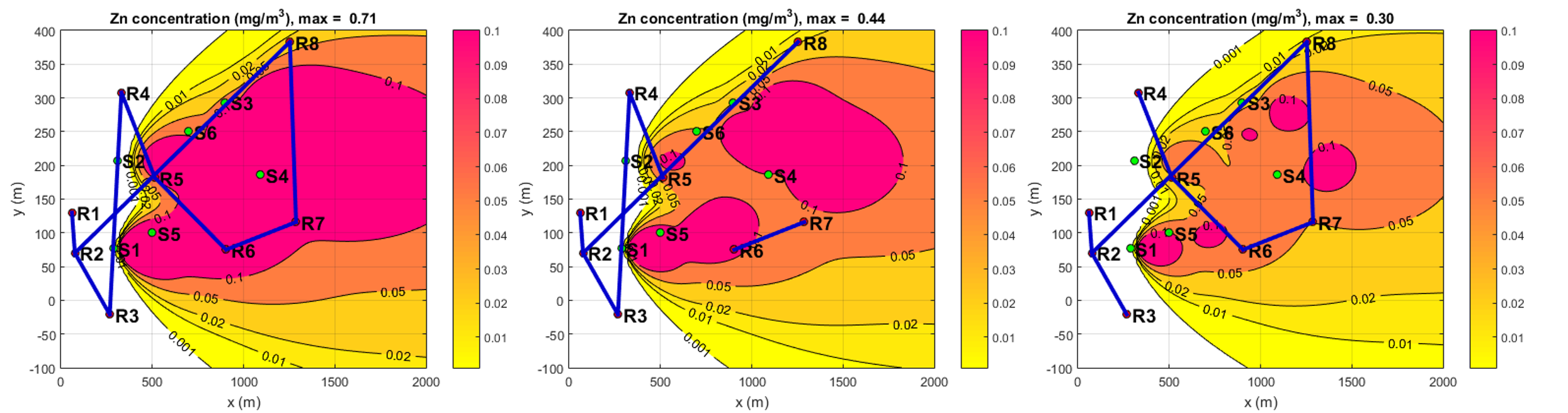}}
	\caption{A motivating example: In an atmospheric dispersion scenario there exists 6 pollutant sources and 8 receptor distributed in the field connected to each other through a time varying graph. At first all receptors are connected and for a time interval we have two disconnected groups. The question is how to handle the consensus over estimates after reconnection. }
	\label{fig:motivating_example}
\end{figure*}
\subsubsection*{Motivating Example}
In fig. \ref{fig:motivating_example} a motivating example is given for the method proposed in this paper. Consider an atmospheric dispersion scenario as an example where there exists 6 pollutant sources and 8 receptor distributed in the field connected to each other through a time varying graph. At first, all receptors are connected  and all the nodes reach a consensus over the field estimate. Later, for a time interval, we have two disconnected groups. The sensors in each group continue receiving new information and calculate their local estimates to the best of their knowledge, After some time the network becomes connected again and agents in each group will get access to the information accumulated in the other group during the disconnection time. As explained earlier, since the priors of the two groups become different, simple averaging is no longer applicable, and using Covariance Intersection results in too conservative estimates. The question is how to handle the consensus over estimates when agents are connected, during the disconnection time, and after reconnection.

In this paper we strive to bring together the best of the two worlds namely, MHMC based distributed averaging, and CI. The former is suitable for reaching consensus over uncorrelated information and the later is useful for combining estimates whose correlations are unknown or difficult to keep track of.  We propose a hybrid scheme that has comparable performance to MHMC consensus while being robust to network failures like the one in Fig. \ref{fig:motivating_example}. Albeit the method is explained with respect to the dynamic field estimation example, it is more generally applicable to most distributed estimation scenarios. 

In Section \ref{subsec:assump_not}, the notation used in this paper is explained as well as assumptions and system model. Section \ref{subsec:Preliminaries} discusses some preliminaries on distributed estimation which paves the way for introducing our problem objective and method. Our proposed method is presented in Section \ref{subsec:Method} along with its theoretical performance analysis compared to MHMC and CI. We extensively evaluate our method's performance in Section \ref{sec:experiments}.  
\section{Modeling} \label{subsec:assump_not}
In this paper the atmospheric dispersion problem is considered as a study case. The three-dimensional advection-diffusion equation describing the contaminant transport in the atmosphere is:
\begin{eqnarray}
\frac{\partial c}{ \partial t} + \nabla \cdot (c \vec{u})  = \nabla \cdot (\vec{K} \nabla c) + Q \delta( \vec{X} - \vec{X_s}),
\end{eqnarray}
where $c(x,y,z,t)$ is the mass concentration at location $\vec{X} = (x, y, z)$, $\vec{X_s}  = (x_s, y_s, z_s)$ is the location of the point source. $\vec{u}$ is the wind velocity, and assume $\vec{u} = (u cos(\alpha), u sin(\alpha), 0) $, for some constant $u \geq 0$, $\alpha$ is the direction of the wind in the horizontal plane and the wind velocity is aligned with the positive $x$-axis when $\alpha = 0$ . $Q$ is the contaminant emitted rate. The eddy diffusivities $\vec{K} = (K_x(x), K_y (x) , K_z (x))$ are functions of the downwind distance $x$ only. Assume that the wind velocity is sufficiently large that the diffusion in the $x$-direction is much smaller than advection, such that the term $K_x \partial_x^2 c$ can be neglected.  The boundary conditions are:
\begin{subequations}\label{eq:boundary_cond}
	\begin{alignat}{2}
& c(0, y, z,t) =  0, c(\infty, y, z,t) =  0,\\
& c(x, \pm \infty, z,t) =  0, c(x, y, \infty,t) =  0, \\
& K_z \frac{\partial c}{\partial z} (x, y, 0,t) =  0.
\end{alignat}
\end{subequations}
With proper discretization of the above PDE, one can define a state vector by stacking the values of the field at a given time $k$ over all sites of the discretization lattice. The PDE model then becomes a lumped parameter, discrete-time linear (LTI) state equation of the form
\begin{equation}
x(k+1)=Ax(k)+Bu(k),
\end{equation}
where $x(k)=\textstyle{[F_k(1,1,1)\cdots F_k(n,n,n)]}$ and $F_k(i_x,i_y,i_z)=c(i_x\Delta_x,i_y\Delta_y,i_z\Delta_z,k\Delta t)$.
\subsubsection*{Stochastic Field Model} Since we consider the case where we have noise and the system is stochastic, we model the evolution of the field using the following equation which relates the state at time step  $k$ to $k+1$: 
\begin{equation}
\label{eq:motion_nodel}
x(k+1)=Ax(k)+Bu(k)+ w(k).
\end{equation}
In the above equation $u(k) \in \mathbb{R}^m $ accounts for $m$ input variables and the vector $w(k)\sim{\cal N}(0, Q(k))$ represents additive white noise used to model unknown perturbations.
\subsubsection*{Network Topology} 
Assume that we have $N$ homogeneous agents associated with nodes of a graph. These agents can communicate with each other under a time-varying network topology $ G_{k} = \langle \mathcal{V}_k,\mathcal{E}_k\rangle $ where $\mathcal{V}_k$ and $\mathcal{E}_k$ are the set of graph nodes and edges respectively. If $(i,j)\in \mathcal{E}_k$, it means agents $i$ and $j$ can communicate. The node corresponding to the $i$-th agent is denoted by $v_i$. Neighbors of node ${v}_i$ are defined as 
\begin{equation}
\mathcal{N}^i = v_i \cup \{ \forall v_j \in \mathcal{V}, (i,j)\in \mathcal{E} \}. 
\end{equation} 
Also $|\mathcal{N}^i|$ is the cardinality of $\mathcal{N}^i$. 

Each agent has a processor and a sensory package on-board. Sensors make observations every $\Delta t$ seconds and processors and the network are fast enough to handle calculations based on message passing among agents every $\delta t$ seconds. We assume that $\delta t \ll \Delta t$. We also assume that the agents exchange their information over the
communication channel which is free of delay or error.

We assume that $x(k)$ denotes the state of the field at time-step $k$. Each agent retains a local version of $x(k)$ which is denoted by ${x}_i(k)$. For random variables we use the following notation: $\hat{x} = \mathbb{E}(x)$ and $P_x =  \mathbb{E}[x-\hat{x}]^2$ are the expected value and the covariance of the random variable $x$ respectively. 
\subsubsection*{Observation Model} We assume that each agent has a sensor that produces noisy observations that are functions of the state of the field. The observation model of the $i$'th  sensor is 
\begin{align}
z_i(k)=H_i(k){x}(k) + v_i(k),\\
v_i(k)\sim{\cal N}(0, R_i(k)).
\end{align}

\section{Decentralized Filtering Preliminaries}
Filtering is the process of recursively computing the posterior probability of a random dynamic process ${x}(k)$ conditioned on a sequence of measurements ${Z}^{k} = \{{z}(1),{z}(2),\dots,{z}(k)\}$, where ${z}(k)$  denotes the observation vector at the time-step $k$. Under the Gaussian assumption, the Kalman Filter (KF) is the optimal  recursive filter for linear state space systems. The KF consists of a prediction and an update. The former uses the motion model to propagate the uncertainty to the next step, and the latter makes adjustments to the predicted values using the most recent observation. We denote the predicted and estimated mean and covariance at time $k$ by $(\hat{x}^-(k),{P}^-(k))$  and $(\hat{x}(k),{P}(k))$.
\label{subsec:Preliminaries}
\subsubsection*{Centralized Kalman Filter}
The KF steps are generally formulated based on the mean and covariance matrix representation of gaussian random variables involved; however, an alternative representation, called the information form of the KF is more useful and intuitive in the development of the decentralized filter. In this representation we define
\begin{subequations}\label{eq:info_rep}
	\begin{alignat}{2}
		{y}(k)&= P_{{{\mathbf{x}}}}^{-1}(k){{\mathbf{x}}}(k), \\
		{Y}(k)&= P_{{{\mathbf{x}}}}^{-1}(k),
	\end{alignat}
\end{subequations} 
where ${y}(k)$ and ${Y}(k)$ are the information vector and information matrix respectively. The prediction step of the KF can then be written as 
\begin{subequations}\label{eq:CIF_pred}
	\begin{alignat}{2}
    	{ M}(k)&={A}^{-T}{Y}(k-1){A}^{-1},\\  
    	P(k)&={ M}(k)+{ Q}(k)^{-1},\\
		{y}^{-}(k)&={ M}(k)-{ M}(k)P(k)^{-1}{ M}(k),  \label{eq:CIF_preda}\\
		{y}^{-}(k)&={Y}^{-}(k){A}{Y}(k-1) {{y}}(k-1).  \label{eq:CIF_predd}
	\end{alignat}
\end{subequations} 
The information content of an observation $z_j(k)$ is ${\delta i}_{j}(k)\displaystyle\mathop{=} {{H}_j^{T}(k)}{{R}_j(k)}^{-1}{z}_j(k)$ along with the information matrix ${\delta I}_{j}(k)\displaystyle\mathop{=}{H}_{j}^T(k){{R}_j(k)}^{-1}{H}_j(k)$. Assuming that information from all agents is available to a central processor, the update step of KF can be carried out by adding the information from different observations to the predicted values. 
\begin{subequations}\label{eq:CIF_update1}
	\begin{alignat}{2}
		{y}(k)&={y}^{-}(k)+\sum\nolimits_{j=1}^{N}{\delta i}_j(k)\\
		{y}(k)&={y}^{-}(k)+\sum\nolimits_{j=1}^{N}{\delta I}_j(k)
	\end{alignat}
\end{subequations} 
This formulation is called the Centralized Information Filter (CIF).

The fundamental assumption in the CIF is that there is a central processor which has access to all the information available. However, when there is no central processor and each agent can only communicate with its neighbors, we want to formulate a decentralized version of the information filter. When run by all agents they should converge to the centralized estimate of the field state.  
\subsection*{Decentralized Estimator Designs}
\subsubsection*{ 1) Consensus Based Estimator}
We start with CIF procedure outlined in previous section. Looking at Eq. \ref{eq:CIF_update1}, one can see that 
${\delta i}(k) \triangleq  N\big(\frac{1}{N} \sum\nolimits_{j=1}^{N}{\delta i}_j(k)\big)\triangleq N\bar{{\delta i}}(k)$ and ${\delta I}(k) \triangleq  N\big(\frac{1}{N} \sum\nolimits_{j=1}^{N}{\delta I}_j(k)\big)\triangleq N\bar{{\delta I}}(k)$. 

Now if all the agents have the same prior information and if via a distributed averaging method the agents can reach a consensus over $\bar{{\delta i}}(k)$ and $\bar{{\delta I}}(k)$, they can use Eq. \ref{eq:CIF_update1} to get a decentralized estimate whose results asymptotically converge to the centralized estimate.

Fortunately, such a method exists. The distributed averaging method of \cite{Boyd2005} makes minimal assumptions about the network topology and only relies on local information exchange between neighboring nodes of a graph to reach a consensus over the average initial value of the nodes. The method uses an iterative linear consensus filter based on the weights calculated from an MHMC. To avoid confusion we use $l$ to indicate consensus iteration throughout this paper. Consider communication graph $ G(l) $.  One can use the message passing protocol of the form $x(l+1)=\textstyle {\sum\nolimits _{j=1}^{|\mathcal{N}_{l}|}} \gamma_{ij}(l)x_{j}(l)$ to calculate the average of the values on the graph nodes in which $d_{i}(l)=|\mathcal{N}^i(l)|$ is the degree of the node $v_i$, and 
\begin{equation}
\label{eq:mhmc}
\gamma_{ij}(l) =
\begin{cases}
{1\over 1+\max\{d_{i}(l),d_{j}(l)\}}      &  \text{if } (i,j) \in \mathcal{E}_{l}, \\
1-\sum_{{(i,m)} \in \mathcal{E}(l)}\gamma_{im}  &  \text{if } i = j,\\
0 & \text{otherwise }.
\end{cases}
\end{equation}
Note that for each node $i$, $\gamma_{ij}$'s only depend on the degrees of its neighboring nodes. Also, due to averaging property of MHMC weights, after reaching consensus, MHMC estimates converge to the centralized estimator's results. Therefore, given the ideal centralized estimate $(\hat{x},P_{x})$, we have $\hat{{x}}_i^{MH}= \hat{{x}}$ and $P_{{{x}}_i}^{MH} = P_{x}$ in the limit.  

In many practical cases the priors become different as a result of network disconnection. In those cases agents have some shared information from the time they were connected to each other and accumulate some new information during the disconnection time. Consequently, priors become different over the network and after reconnection, their consensus should be handled carefully. 

\subsubsection*{2) Covariance Intersection Based Estimator} It follows from the above discussion that if the priors are not the same among the network nodes, distributed averaging alone will not produce consistent estimates. One way of handling such a scenario is using Covariance Intersection (CI) methods. We may use an iterative Covariance Intersection method to reach a consensus over the local estimates when the priors are not the same, either due to disconnection or early stopping of consensus process. In iterative CI, the goal is to fuse different estimates of a random variable without having any knowledge about the cross covariance between such estimates. Iterative CI, iteratively solves the following optimization problem and updates local estimates accordingly until it reaches consensus. 
 \subsubsection*{Iterative CI} At each iteration $l$, for each agent, define the local information to be  $$\mathcal{I}_i(l) \displaystyle\mathop{=}^{\Delta} {Y}_{i}(l) +{\delta I_i} (l).$$ 
 Solve for $w^*$ such that 
 \begin{equation}
 \label{eq:opt_problem}
   \begin{aligned}
  \omega^* &= { \argmin_{\omega}} \mathcal{J} \big( [{\sum\nolimits_{j \in {\cal N}^{i}(l)}\omega_{j} }\mathcal{I}_j(l)]^{-1} \big), \\
  s.t. &\quad  {\sum\nolimits_{j=1}^{|{\cal N}^{i} (l)|}} \omega_{j}= 1, \quad \forall j \quad \omega_j\geq 0,
   \end{aligned}
    \end{equation}
where $\mathcal{J}(\cdot)$ is an optimization objective function; It can be $trace(\cdot)$ or $\log\det(\cdot)$. Estimates are then updated locally for the next iteration    
 \begin{align}
  {{y_i}(l+1)}&=\textstyle {\sum_{j \in {\cal N}^{i}(l)}} \omega_{j}^* [{y}_{j}(l) + {\delta i_j} (l)], \\{{Y_i}(l+1)}&=\textstyle {\sum_{j \in {\cal N}^{i}(l)}}\omega_{j}^* [{Y}_{j}(l) +{\delta I_j} (l)].
  \end{align}
As will be discussed in Section \ref{subsec:Method}, CI and iterative CI (ICI) generate conservative estimates which means that $	\mathbb{E}[x-\hat{{x}}_i^{CI}] =	\mathbb{E}[x-\hat{{x}}_i]=0$ and $P_{{{x}}_i}^{CI} \geq P_{x}$ for the local estimates and the consensus value. The disadvantage of CI is that it generates overly conservative estimates by continually neglecting the cross correlation information. 
\label{mot_exmpl}
\begin{algorithm}[!t]
\label{alg:inf-predict}
\SetKwInOut{Input}{Input}
    \SetKwInOut{Output}{Output}
      \Input{$[{y}_j(t_0),{Y}_j(t_0)]$ }
\caption{Proposed Method}
Use Eqns. \ref{eq:CIF_preda}  -  \ref{eq:CIF_predd} to calculate predicted values $[{y}_j^-(t_1) ,{Y}_j^-(t_1)]$ given $[{y}_j(t_0),{Y}_j(t_0)]$
Collect local observation ${z}_j(t_1)$ and calculate jacobian and noise covariance  $[{H}_j(t_1),{R}_j(t_1)]$ 
Calculate the local information update
\begin{align*}
	{\delta i_j} (t_1) &= {H}_{j}^{T}(t_1){R}_j^{-1}(t_1){z}_j(t_1)\\
	{\delta I_j} (t_1) &= {H}_{j}^{T}(t_1){R}_j^{-1}(t_1){H}_{j}(t_1) 
\end{align*}
Initialize consensus variables ($l=0$)
\begin{align*}
	[{ {\mathpzc{y}_j^{k}}},{ \mathpzc{Y}_j^{l}}]&=[{y}_j^-(t_1) ,{Y}_j^-(t_1)]\\
    [{{\mathpzc{\overline {\delta i}}_j^{l}}},{ \mathpzc{\overline {\delta I}}_j^{l}}]&=[{\delta i}_j(t_1),{\delta I}_j(t_1)]
\end{align*}
 \While{\texttt{NOT COVERGED}} 
    { $\texttt{BROADCAST}[ \mathpzc{y}_j^{l}, \mathpzc{Y}_j^{l},{{\mathpzc{\overline{\delta i}}_j^{l}}},{ \mathpzc{\overline{\delta I}}_j^{l}}]$\\
     $\texttt{RECEIVE}[ \mathpzc{y}_l^{l}, \mathpzc{Y}_l^{l},{{\mathpzc{\overline{\delta i}}_l^{l}}},{ \mathpzc{\overline{\delta I}}_l^{l}}] \quad \forall l \in \mathcal{N}^j(l)$ \\
     Collect received data 
\begin{align*}
   \mathcal{C}_j^k=\{{ {\mathpzc{y}_l^k}},{ \mathpzc{Y}_l^k} \mid l\in \mathcal{N}^j(k) \}\\
     \mathcal{M}_j^k= \{{{\mathpzc{\overline{\delta i}}_l^k}},{ \mathpzc{\overline{\delta I}}_l^k} \mid l\in \mathcal{N}^j(k)\}
\end{align*}
Do one iteration of CI on consensus variables for local prior information $\mathcal{C}_j^l$
$$[{ {\mathpzc{y}_j^{l+1}}},{ \mathpzc{Y}_j^{l+1}}]=\texttt{CI}(\mathcal{C}_j^l)$$\\
Do one iteration of MHMC on consensus variables for new information $\mathcal{C}_j^l$
$$[{{\mathpzc{\overline{\delta i}}_j^{l+1}}},{ \mathpzc{\overline{\delta I}}_j^{l+1}}]=\texttt{MHMC}(\mathcal{M}_j^l)$$\\
$l \leftarrow  l+1$
   }
Calculate the posteriors according to:
\begin{align*}
{Y}_j(t_1) &= { {\mathpzc{Y}_j^{l}}} + n_{CG}{{\mathpzc{\overline{\delta I}}_j^{l}}} \\
{y}_j(t_1) &= { {\mathpzc{y}_j^{l}}} + n_{CG}{{\mathpzc{\overline{\delta i}}_j^{l}}}
\end{align*}
\Return ${Y}_j(t_1),{y}_j(t_1)] $
\end{algorithm}

\subsection*{Problem Objective}

Our goal is to design a network agnostic recursive decentralized estimator to calculate the local estimate ${x}_i$ along with an associated covariance $P_{x_i}$such that the following properties hold: 
\label{eq:objective_fun}
	\begin{align}
	\label{eq:objective_fun1}
&	\mathbb{E}[x-\hat{{x}}_i^{CI}] =	\mathbb{E}[x-\hat{{x}}_i] =	\mathbb{E}[x-\hat{{x}}_i^{MH}]=0, \nonumber \\
	&\mathcal{J}( P_{{{x}}_i}^{MH})\leq \mathcal{J}(P_{{{x}}_i}) \leq \mathcal{J}(P_{{{x}}_i}^{CI}), 
	\end{align}
i.e., we are looking for an unbiased estimate whose covariance is less than that of CI. 
\section{Hybrid CI consensus}

\label{subsec:Method}
We propose to use iterative CI to reach consensus over priors and the MHMC based consensus filter for distributed averaging of local information updates. Our method is summarized in Algorithm \ref{alg:inf-predict}. We explain the flow of the proposed method using a simple scenario with  two agents. Generalization to more than two agents is straightforward and follows similar steps. 

Imagine a scenario consisting of two agents, observing a dynamic field with state vector $x$, that are communicating with each other through a time-varying network topology. At time $t_0$, the agents start with  priors $[{y}_1^-(t_0) ,{Y}_1^-(t_0)]$ and $[{y}_2^-(t_0) ,{Y}_2^-(t_0)]$ respectively.

At time $t_1$ the field evolves to the new state $x(t_1)$ and agents calculate their own local prediction (line 1 in the algorithm). Then they make observations $z_1(t_1)$ and $z_2(t_1)$, respectively, and compute the local information updates $[{\delta i}_{1}(t_1),{\delta I}_{1}(t_1)]$ and $[{\delta i}_{2}(t_1),{\delta I}_{2}(t_1)]$ (lines 2 and 3 of the algorithm). 

The two agents, if performing CI, would find a fused estimate $$ { {{Y^{CI}}}} = w^{CI} ({Y}_1^- + {\delta I}_{1})  + (1-w^{CI})({Y}_2^-+ {\delta I}_{1}),$$
where $w^{CI}$ is obtained from solving the optimization problem in Eq. \ref{eq:opt_problem}. Note that doing MHMC alone is not possible here since ${Y}_1^-$ and ${Y}_2^-$ are different. In our hybrid method we do the following:
$$Y^{Hyb} =\underbrace{w^{Hyb} {Y}_1^- + (1-w^{Hyb}){Y}_2^-}_{  \substack{\text{CI to reach }\\ \text{ consensus over priors}}}    +\underbrace{ {\delta I}_{1} + {\delta I}_{2}}_{\substack{\text{consensus over} \\ \text{the  incremental } \\ \text{information}   }}.$$
It can be seen that  ${\delta I}_{1} + {\delta I}_{2}\geq w^{CI}{\delta I}_{1} + (1-w^{CI}){\delta I}_{2}$
and $\mathcal{J}(w^{Hyb} {Y}_1^- + (1-w^{Hyb}){Y}_2^-) \geq \mathcal{J}(w^{CI} {Y}_1^- + (1-w^{CI}){Y}_2^-)$
due to the fact that optimization problem for ${Y}_2^-$ and ${Y}_2^-$ has the optima  $w^{Hyb}$. If $\mathcal{J}(\cdot)$ has the property that if  $
\mathcal{J}(\mathcal{Y}_1) \geq \mathcal{J}(\mathcal{Y}_2)$ and $\mathcal{I}_1 \geq \mathcal{I}_2$ then $\mathcal{J}(\mathcal{Y}_1 + \mathcal{I}_1) \ge \mathcal{J}(\mathcal{Y}_2 + \mathcal{I}_2)$, then our method is guaranteed to outperform CI. 

For an $N$-agent system with the $i'th$ agent having prior $Y_i^-$, the ICI approach is used to find a consensus over the priors using Eq. \ref{eq:opt_problem} recursively. The MHMC approach is used to form the consensus over the new information, i.e., $\textstyle \sum\nolimits_{j=1}^{N}{\delta I}_j$ (Eq. \ref{eq:mhmc}). In line $12$ of the algorithm, $n_{CG}$ is the number of agents that form a connected group, and it can be determined by assigning unique IDs to the agents and passing these IDs along with the consensus variables. Each agent keeps track of unique IDs it receives and passes them to its neighbors.
The following  propositions hold.
\prop{If the objective function $\mathcal{J}$ is strictly convex, the ICI process is guaranteed to reach a consensus over the priors, i.e., $\exists Y_{\infty}^-$, $Y_{i}^- (l)\rightarrow Y_{\infty}^- \quad \forall i$ as $l\rightarrow \infty$.} 
The same result holds for the information vector as well. 
\begin{proof}
At each iteration '$l$' and for each agent '$j$', ICI solves an instance of the optimization problem in Eq. \ref{eq:opt_problem}. The ICI updates local variables, ${Y}_i(l)$, according to
 \begin{equation}
  {{Y}_i(l+1)}= {\sum_{j \in {\cal N}^{i}(l)}} \omega_{j}^* {Y}_j(l).
  \end{equation}
The very definition of the optimization problem requires that\footnote{Can be easily proved by contradiction.}
 \begin{equation}
 \label{eq:ICI1}
\mathcal{J}( {{Y}_i^{-1}(l+1)}) \le \mathcal{J}( {{Y}_j^{-1}(l)}) \quad \forall j \in {\cal N}^{i}(l)
 \end{equation}
Lets define $V({Y}_i,l) = \mathcal{J}( {{Y}_i^{-1}(l)})$. Take the Lyapunov function of the whole network at iteration $l$ to be 
\begin{equation}
\mathcal{V}({Y}_1,\cdots {Y}_N,l) = \sum\limits_{i=1}^{N}V({Y}_i,l).
\end{equation}
If $\mathcal{J}(X^{-1})$ is a positive and monotonically decreasing function over the set of $\{X \in   \mathcal{S}^+ \triangleq $ Symmetric Positive Definite matrices$\}$, i.e., $\mathcal{J}(X^{-1})>0\quad \forall X \in   \mathcal{S}^+$ and 
$$\forall X,Y \in \mathcal{S}^+, \quad X\ge Y \Rightarrow \mathcal{J}(X^{-1})\leq \mathcal{J}(Y^{-1}),$$ then  $ \forall l,\quad \mathcal{V}({Y}_1,\cdots {Y}_N,l)>0$. Also, based on Eq. \ref{eq:ICI1} $\mathcal{V}({Y}_1,\cdots {Y}_N,l+1) \leq \mathcal{V}({Y}_1,\cdots {Y}_N,l)$. Since $\mathcal{V}({Y}_1,\cdots {Y}_N,l)$ is monotonically decreasing and since it is a positive function, it has a lower bound $>0$. If it reaches this lower bound, all ${Y}_i$'s should be equal; otherwise, $\mathcal{V}({Y}_1,\cdots {Y}_N,l+1) < \mathcal{V}({Y}_1,\cdots {Y}_N,l) $ due to the strictly convex property of $\mathcal{J}$ which is a contradiction. 
Therefore, by performing ICI, the Lyapunov function of the network is guaranteed to reach a lower bound in which all ${Y}_i$'s are equal. Therefore if there exists a ${Y}_{\infty}$ then the network is guaranteed to converge to it. Conditions for $\mathcal{J}(X^{-1})$ are satisfied for $\mathcal{J}(X^{-1}) = \log \det (X^{-1})$. $\qed$
\end{proof}
\remark{ It is straghtforward to show that ICI and our method both produce unbiased estimates. Regardig the second part of Eq. \ref{eq:objective_fun1}, our extensive experiments on the system considered in this paper and various other systems show that the inequality holds for $\mathcal{J}=\log\det(\cdot)$. The existence of a theoretical proof is currently being investigated and will be the subject of future work.}
\normalfont

\section{Experiments} \label{sec:experiments}
We perform two sets of experiments on an atmospheric dispersion problem to show the effectiveness of our method and evaluate its performance during disconnection and after reconnection. This is a three dimensional problem and after proper discretizing of its Partial Differential Equation (PDE), we get a system in the format of Eq. \ref{eq:motion_nodel}. 

For our experiments after discretization, the dimension of the state is 80. We assume that there are 10 sources emitting pollutant Zinc (referred to as $Zn$ from now on) into atmosphere. There are also 9 receptors making noisy measurements of the concentrations of $Zn$ around their location in space. We assume that receptors can communicate to each other through a time varying network which does not remain  connected at all times. Receptors receive information only from their immediate neighbors. They all have access to the sources' locations and the source emission is modeled as a white noise process with known covariance. 

\subsubsection{Investigating the effect of disconnection on estimation performance}
\begin{figure}[t]
		\centering
		{\includegraphics[width=7.5cm]{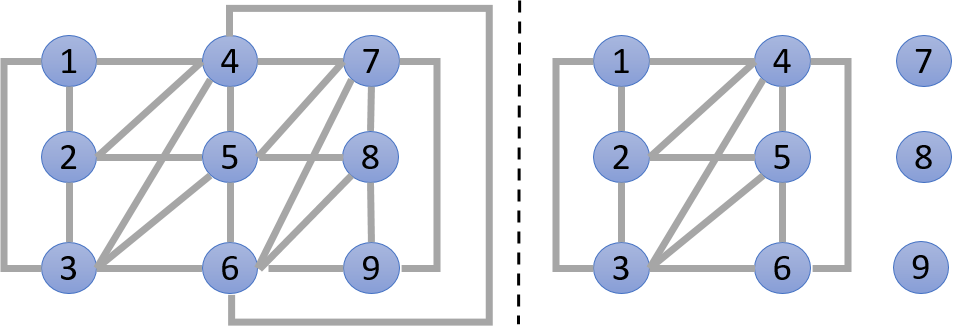}}
		\caption{ Topology of the Network when all receptors are connected (left) and when receptors 7,8 and 9 get disconnected from the rest of the group (right). }
		\label{fig:top}
	\end{figure}

In this experiment we intend to evaluate the performance of the proposed method during the phase where some receptors become disconnected from the rest of the group and get connected again after some interval. The topology of the network takes one of the forms depicted in Fig. \ref{fig:top}. The network starts fully connected and starting from timestep 3, receptors 7, 8 and 9 become isolated and remain in this situation for 2 steps, then they are connected back to the rest of the receptors. Similarly, disconnection happens in intervals $[17 - 20]$ and $[23 - 30]$.      

\begin{figure}[t]
	\resizebox{16cm}{15cm}{{\input{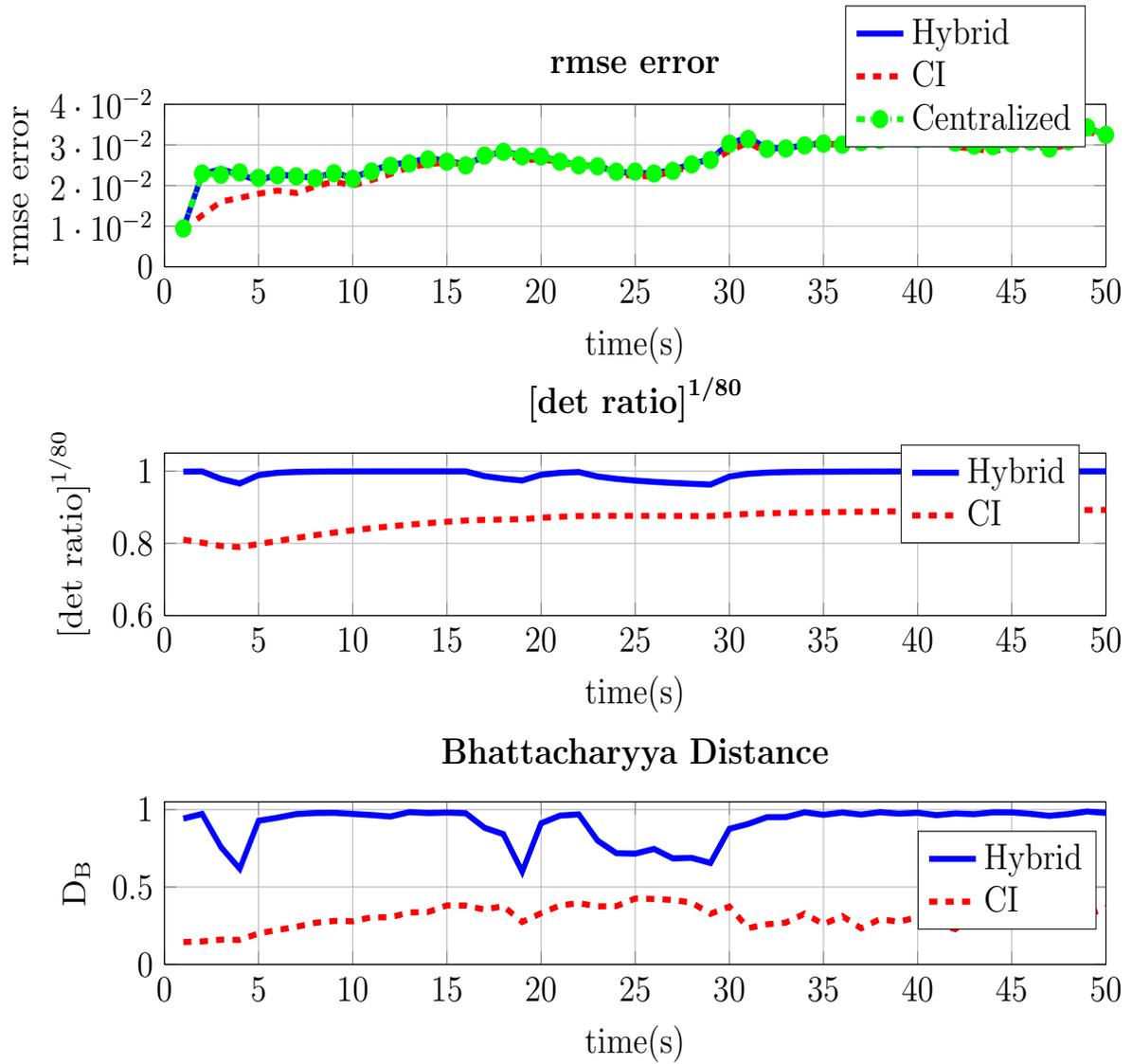} }}
		\caption{Comparison of the estimation results using centralized kalman filter, pure covariance intersection, and our method. }
		\label{fig:sim_error_comp}
	\end{figure}

In order to make a comparison we obtain the estimation result using pure CI, our method and also a centralized estimator to see how much of its performance can be recovered. Note that the MHMC consensus cannot be done here due to disconnection. The results are depicted in Fig. \ref{fig:sim_error_comp}.
We use three measures to evaluate the estimates.
\begin{enumerate}
\item The Bhattacharyya distance \cite{bhattacharyya1946measure} between the estimation  results and the centralized estimator. The Bhattacharyya distance can be used to evaluate the similarity of two continuous probability distribution function. For gaussian distributions parametrized as $ (\mu_1,\Sigma_1)$, and  $(\mu_2,\Sigma_2)$, $0 \leq D_{B}(p_1,p_2)\leq 1$ and is defined:
\begin{align}
 & D_{B}(p_1,p_2) = e^{-D(p_1,p_2)}\\
 & D(p_1,p_2) = \frac{1}{8}(\mu_1 - \mu_2)^T\Sigma(\mu_1 - \mu_2)+\nonumber \\
 &\frac{1}{2}\ln\frac{\det \Sigma}{\sqrt{\det \Sigma_1 \det \Sigma_2}}, \quad \Sigma = \frac{\Sigma_1+\Sigma_2}{2} \nonumber
\end{align}
Here $D_{B}(p_1,p_2)=1$ means complete similarity and $D_{B}(p_1,p_2)=0$ means complete dissimilarity.
\item $[\frac{\det P_{cen}}{\det P_{CI/Hyb}}]^{\frac{1}{n_d}}$ which is a non-dimensionalized measure of covariance volume ratio in which ${n_d}$ is the dimensionality of the state vector. \item $rmse$ estimation error.
\end{enumerate}

As it can be seen, the proposed method outperforms pure CI as expected and is able to get the performance very close to centralized estimator results after reconnection.  Based on Bhattacharyya distance, closeness between centralized and decentralized estimators drops during disconnection interval as expected since receptors do not have access to all the information available to the centralized estimator. While the proposed method is able to immediately recover after reconnection, pure CI continues to have lower performance even after re-connection due to the fact that it ignores the correlations. 
	 		\begin{figure}[t]
	 			\centering
	 			{\includegraphics[width=16cm]{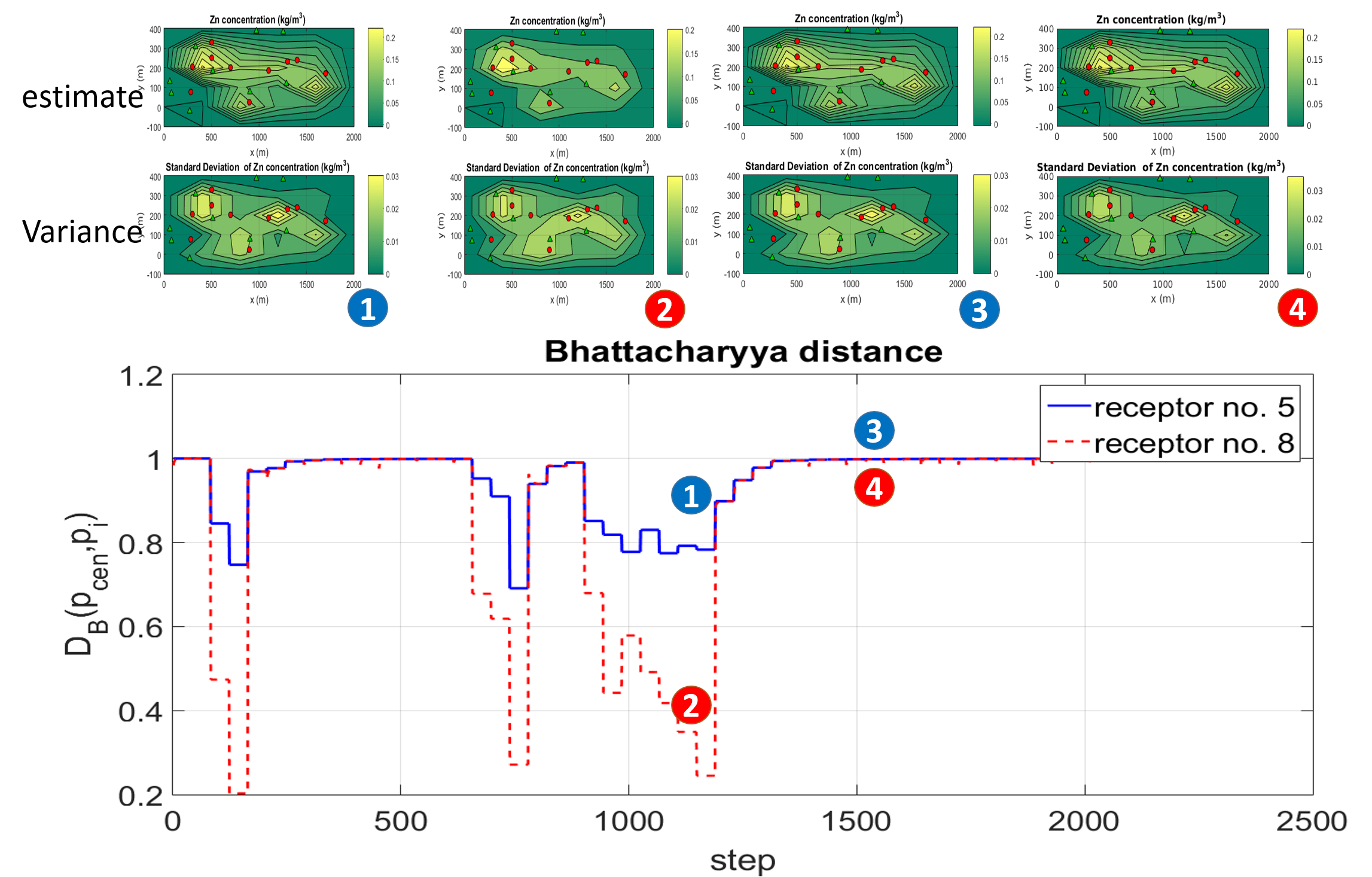}}
	 			\caption{ Estimation performance comparison among receptors.}
	 			\label{fig:sim_rec}
	 		\end{figure}

	 			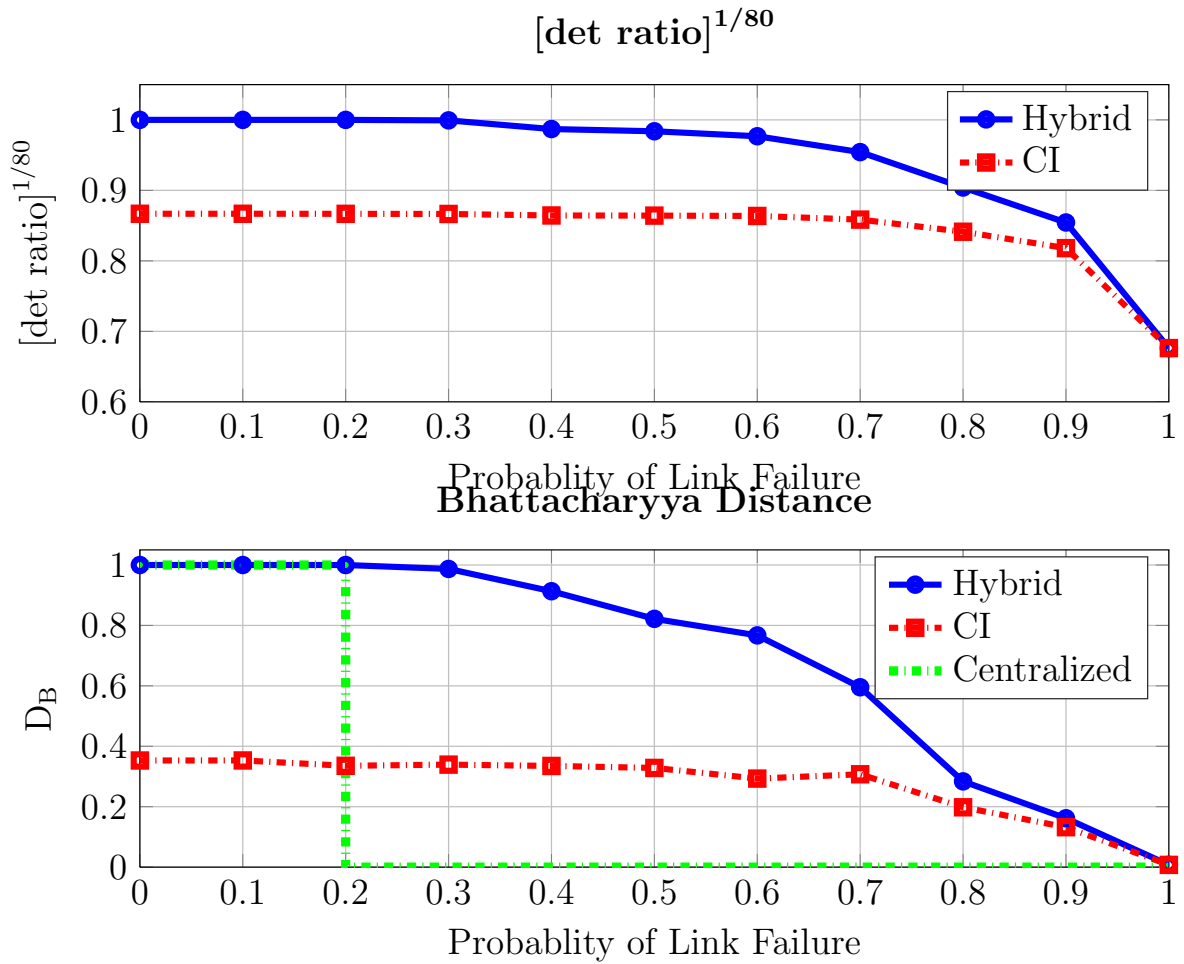
\begin{figure}[t]
	 					\centering
	 						\resizebox{16cm}{13cm}{{
%
%
\begin{tikzpicture}

\begin{axis}[%
width=0.777\textwidth,
height=0.24\textwidth,
at={(0\textwidth,0.352\textwidth)},
scale only axis,
xmin=0,
xmax=1,
xtick={  0, 0.1, 0.2, 0.3, 0.4, 0.5, 0.6, 0.7, 0.8, 0.9,   1},
xlabel={Probablity of Link Failure},
xmajorgrids,
ymin=0.6,
ymax=1.05,
ytick={0.6, 0.7, 0.8, 0.9,   1},
ylabel={${\text{[det ratio]}}^{\text{1/80}}$},
ymajorgrids,
axis background/.style={fill=white},
title style={font=\bfseries},
title={${\text{[det ratio]}}^{\text{1/80}}$},
legend style={legend cell align=left,align=left,draw=white!15!black}
]
\addplot [color=blue,solid,line width=2.0pt,mark=o,mark options={solid}]
  table[row sep=crcr]{%
0	0.999999999999967\\
0.1	1.00000158269278\\
0.2	1.00000212926993\\
0.3	0.999290618288621\\
0.4	0.986990187832471\\
0.5	0.983829676647235\\
0.6	0.976770402996627\\
0.7	0.954276103799111\\
0.8	0.904081387256995\\
0.9	0.854431085355276\\
1	0.676275176380441\\
};
\addlegendentry{Hybrid};

\addplot [color=red,dashdotted,line width=2.0pt,mark=square,mark options={solid}]
  table[row sep=crcr]{%
0	0.866829634767757\\
0.1	0.86681620843843\\
0.2	0.866723664183391\\
0.3	0.866542605167923\\
0.4	0.864403910850494\\
0.5	0.864197109727685\\
0.6	0.863707357291093\\
0.7	0.858520270746827\\
0.8	0.841323297941579\\
0.9	0.818059487271007\\
1	0.676275176380439\\
};
\addlegendentry{CI};

\end{axis}

\begin{axis}[%
width=0.777\textwidth,
height=0.24\textwidth,
at={(0\textwidth,0\textwidth)},
scale only axis,
xmin=0,
xmax=1,
xtick={  0, 0.1, 0.2, 0.3, 0.4, 0.5, 0.6, 0.7, 0.8, 0.9,   1},
xlabel={Probablity of Link Failure},
xmajorgrids,
ymin=0,
ymax=1.05,
ytick={  0, 0.2, 0.4, 0.6, 0.8,   1},
ylabel={$\text{D}_{\text{B}}$},
ymajorgrids,
axis background/.style={fill=white},
title style={font=\bfseries},
title={Bhattacharyya Distance},
legend style={legend cell align=left,align=left,draw=white!15!black}
]
\addplot [color=blue,solid,line width=2.0pt,mark=o,mark options={solid}]
  table[row sep=crcr]{%
0	0.999999999999997\\
0.1	0.999998716429475\\
0.2	0.99997984768421\\
0.3	0.987029616939263\\
0.4	0.913185617584752\\
0.5	0.821814186191355\\
0.6	0.767002829655032\\
0.7	0.595338072485605\\
0.8	0.283817946464577\\
0.9	0.161705718471468\\
1	0.00748472330063145\\
};
\addlegendentry{Hybrid};

\addplot [color=red,dashdotted,line width=2.0pt,mark=square,mark options={solid}]
  table[row sep=crcr]{%
0	0.352940661300763\\
0.1	0.353081070206302\\
0.2	0.335426196995786\\
0.3	0.339269640535606\\
0.4	0.334930218472385\\
0.5	0.328360914514182\\
0.6	0.293069210645253\\
0.7	0.307328795394825\\
0.8	0.198078002628855\\
0.9	0.132169488776635\\
1	0.00748472330062795\\
};
\addlegendentry{CI};

\addplot [color=green,dashdotted,line width=3.0pt]
  table[row sep=crcr]{%
0	1\\
0.1	1\\
0.2	1\\
0.2	0\\
0.3	0\\
0.4	0\\
0.5	0\\
0.6	0\\
0.7	0\\
0.8	0\\
0.9	0\\
1	0\\
};
\addlegendentry{Centralized};

\end{axis}
\end{tikzpicture}%
 }}

	 					\caption{Composite diagram for performance comparison for different probablity of link failure. }
	 					\label{fig:simfail}
	 				\end{figure}
Fig. \ref{fig:sim_rec} takes a closer look at the performance of the proposed method and compares the estimation results of receptor 5 and 8 during two different time steps. The vertical axes represent consensus steps not time. Based on Fig.  \ref{fig:top}, receptor 5 remains in a group of size 6 during disconnection period whereas receptor 8 remains totally isolated in that period. The higher difference between centralized and decentralized estimate for this receptor can be explained based on the fact that it has less information at its disposal. However, after reconnection both receptors are able to converge to the same value which is very close to the centralized estimator. 
	 		\subsubsection{Performance analysis and robustness to link failure} 
In this experiment we evaluate the performance of our method in a systematic way to establish its usefulness and robustness to networks with high probability of link failure. We consider the same system as in the first experiment and simulate it for 50 time steps. At the beginning of each step a 4 regular graph with 9 nodes is generated, and given a probability of failure for each link, some links in the graph will randomly be disconnected. Depending on the regularity degree, and probably of failure, in some percentage of times, the graph still remains connected. However, if the regularity degree goes down or the probability of failure increases, more often than not, the graph becomes disconnected. 

In practice, for $p\geq0.2$, consensus methods are no longer guaranteed to succeed since the network almost always get disconnected at some point in time.

 We ran our method for 50 steps, as explained earlier, for each probability of link failure and compared its performance with the ideal centralized result (which is obtained by assuming full connectivity at all times). The performance is evaluated by calculating the average value for Bhattacharyya distance and determinant ratio measure at all steps and for all receptors.  Based on Fig. \ref{fig:simfail}, for the case considered in this experiment, our decentralized estimator performs very similar to the ideal centralized one for $p\in[0.0 ,0.4]$ while drastically outperforming pure CI all the time. This means that in the case considered here, our method can perform almost as well as the ideal estimator for an unreliable  network. Obviously, the performance can vary from one system to another and under different network topologies. However, our method can recover the performance of the centralized method when the network is unreliable and substantially outperforms pure CI always as it has already been established theoretically. 
	\section{Conclusion} \label{sec:conclusion}
	This paper proposes a decentralized estimator for dynamic systems in networks with changing topology and those that do not remain connected all the time. Separating the process of consensus for the correlated and uncorrelated information was the key to achieve a better performance compared to Covariance Intersection alone. Evaluating the proposed method on an $80$-dimensional estimation problem showed substantial performance improvement compared to CI and also the ability to recover after a disconnection interval occurs.

	\bibliographystyle{IEEEtran} 
	\bibliography{technical_report}

\begin{thebibliography}{2}
\providecommand{\natexlab}[1]{#1}
\providecommand{\url}[1]{\texttt{#1}}
\expandafter\ifx\csname urlstyle\endcsname\relax
  \providecommand{\doi}[1]{doi: #1}\else
  \providecommand{\doi}{doi: \begingroup \urlstyle{rm}\Url}\fi

\bibitem[Kalman(1960)]{kalman1960new}
R.E. Kalman.
\newblock A new approach to linear filtering and prediction problems.
\newblock \emph{Journal of Basic Engineering}, 82\penalty0 (1):\penalty0
  35--45, 1960.

\bibitem[McGeer(1990)]{McGeer01041990}
Tad McGeer.
\newblock \href{http://ijr.sagepub.com/content/9/2/62.abstract}{Passive Dynamic
  Walking}.
\newblock \emph{The International Journal of Robotics Research}, 9\penalty0
  (2):\penalty0 62--82, 1990.
\newblock \doi{10.1177/027836499000900206}.
\newblock URL \url{http://ijr.sagepub.com/content/9/2/62.abstract}.

\end{thebibliography}


\begin{thebibliography}{10}
\providecommand{\url}[1]{#1}
\csname url@rmstyle\endcsname
\providecommand{\newblock}{\relax}
\providecommand{\bibinfo}[2]{#2}
\providecommand\BIBentrySTDinterwordspacing{\spaceskip=0pt\relax}
\providecommand\BIBentryALTinterwordstretchfactor{4}
\providecommand\BIBentryALTinterwordspacing{\spaceskip=\fontdimen2\font plus
\BIBentryALTinterwordstretchfactor\fontdimen3\font minus
  \fontdimen4\font\relax}
\providecommand\BIBforeignlanguage[2]{{%
\expandafter\ifx\csname l@#1\endcsname\relax
\typeout{** WARNING: IEEEtran.bst: No hyphenation pattern has been}%
\typeout{** loaded for the language `#1'. Using the pattern for}%
\typeout{** the default language instead.}%
\else
\language=\csname l@#1\endcsname
\fi
#2}}

\bibitem{Zhang_ttradeof1}
H.~Zhang, J.~Moura, and B.~Krogh, ``Dynamic field estimation using wireless
  sensor networks: Tradeoffs between estimation error and communication cost,''
  \emph{Signal Processing, IEEE Transactions on}, vol.~57, no.~6, pp.
  2383--2395, June 2009.

\bibitem{olfati2005distributed}
R.~Olfati-Saber, ``Distributed kalman filter with embedded consensus filters,''
  in \emph{Decision and Control, 2005 and 2005 European Control Conference.
  CDC-ECC'05. 44th IEEE Conference on}.\hskip 1em plus 0.5em minus 0.4em\relax
  IEEE, 2005, pp. 8179--8184.

\bibitem{olfati2005consensus}
R.~Olfati-Saber and J.~S. Shamma, ``Consensus filters for sensor networks and
  distributed sensor fusion,'' in \emph{Decision and Control, 2005 and 2005
  European Control Conference. CDC-ECC'05. 44th IEEE Conference on}.\hskip 1em
  plus 0.5em minus 0.4em\relax IEEE, 2005, pp. 6698--6703.

\bibitem{Boyd2005}
L.~Xiao, S.~Boyd, and S.~Lall, ``A scheme for robust distributed sensor fusion
  based on average consensus,'' in \emph{Information Processing in Sensor
  Networks, 2005. IPSN 2005. Fourth International Symposium on}, April 2005,
  pp. 63--70.

\bibitem{capitan_delayed_icra_2009}
J.~Capitan, L.~Merino, F.~Caballero, and A.~Ollero, ``Delayed-state information
  filter for cooperative decentralized tracking,'' in \emph{Robotics and
  Automation, 2009. ICRA '09. IEEE International Conference on}, May 2009, pp.
  3865--3870.

\bibitem{chen2002estimation}
L.~Chen, P.~O. Arambel, and R.~K. Mehra, ``Estimation under unknown
  correlation: covariance intersection revisited,'' \emph{Automatic Control,
  IEEE Transactions on}, vol.~47, no.~11, pp. 1879--1882, 2002.

\bibitem{wang_distr_CI}
Y.~Wang and X.~Li, ``Distributed estimation fusion with unavailable
  cross-correlation,'' \emph{Aerospace and Electronic Systems, IEEE
  Transactions on}, vol.~48, no.~1, pp. 259--278, Jan 2012.

\bibitem{dist_inf_filter_2009}
D.~Casbeer and R.~Beard, ``Distributed information filtering using consensus
  filters,'' in \emph{American Control Conference, 2009. ACC '09.}, June 2009,
  pp. 1882--1887.

\bibitem{hu2012diffusion}
J.~Hu, L.~Xie, and C.~Zhang, ``Diffusion kalman filtering based on covariance
  intersection,'' \emph{Signal Processing, IEEE Transactions on}, vol.~60,
  no.~2, pp. 891--902, 2012.

\bibitem{deng2012sequential}
Z.~Deng, P.~Zhang, W.~Qi, J.~Liu, and Y.~Gao, ``Sequential covariance
  intersection fusion kalman filter,'' \emph{Information Sciences}, vol. 189,
  pp. 293--309, 2012.

\bibitem{cristofaro2013distributed}
A.~Cristofaro, A.~Renzaglia, and A.~Martinelli, ``Distributed information
  filters for mav cooperative localization,'' in \emph{Distributed Autonomous
  Robotic Systems}.\hskip 1em plus 0.5em minus 0.4em\relax Springer, 2013, pp.
  133--146.

\bibitem{bhattacharyya1946measure}
A.~Bhattacharyya, ``On a measure of divergence between two multinomial
  populations,'' \emph{Sankhy{\=a}: The Indian Journal of Statistics}, pp.
  401--406, 1946.

\end{thebibliography}
	
	
\end{document}